\documentclass[apjl]{emulateapj}
\slugcomment{Not to appear in Nonlearned J., 45.}

\shorttitle{}
\shortauthors{}

\begin{document}


\title{First evidence for dusty disks around Herbig Be stars}

\author{A. Fuente\altaffilmark{1},A. Rodr\'{\i}guez-Franco\altaffilmark{2},
L. Testi\altaffilmark{3}, A. Natta\altaffilmark{3}, R. Bachiller\altaffilmark{1},
R. Neri\altaffilmark{4}}
\altaffiltext{1}{Observatorio Astron\'omico Nacional, Apdo. 1143, E-28800 Alcal\'a de Henares,Spain;
a.fuente@oan.es}
\altaffiltext{2}{Dpto Matem\'atica Aplicada, Universidad Complutense de Madrid, Av. Arcos de Jal\'on s/n, E-28037 Madrid, Spain}
\altaffiltext{3}{Osservatorio Astrofisico di Arcetri, Largo Enrico Fermi, 5, I-50125 Firenze, Italy}
\altaffiltext{4}{Institute de Radioastronomie Millim\'etrique, 300 rue de la Piscine, 38406 St Martin d'Heres Cedex, France}
%\altaffiltext{5}{Instituto de Estructura de la Materia, CSIC, Serrano 121, E-28006 Madrid, Spain}


\begin{abstract}
We have carried out a high-sensitivity search for circumstellar disks around
Herbig Be stars in the continuum at 1.4mm and 2.7mm using the IRAM interferometer
at the Plateau de Bure (PdBI) . In this letter, we report data on  three
well studied B0 stars, MWC 1080, MWC 137 and R Mon. 
The two latter have also been observed in the continuum at 0.7 cm and 1.3 cm 
using the NRAO Very Large Array (VLA) . 
We report the detection of 
circumstellar disks around MWC 1080 and R Mon with  masses of
M$_d \sim$ 0.003 and  0.01 M$_\odot$, respectively, while for
MWC 137 we estimate a disk mass upper limit of 0.007 M$_\odot$. 
Our results show that
the ratio M$_d$/M$_*$ is at least an order of magnitude lower in Herbig Be stars than in 
Herbig Ae and  T Tauri stars.
\end{abstract}




\section{Introduction}
The existence of accretion disks around massive stars (M$>$ 5 M$_\odot$) 
remains a matter of debate. There is increasing evidence for
the existence of flattened structures (disks) around high-mass protostars.
However there is no clear evidence of disks in later phases,
namely in Herbig Be (HBe) stars.  
 \citet{nat00} compiled the interferometric
observations in mm continuum around HBe stars and found that 
the occurrence of disks in HBe is $\sim$ 0. This contrasts with
the case of the lower mass HAe stars (2M$_\odot$ $\leq$ M$_*$ $\leq$ 5 M$_\odot$) 
in which the occurrence of circumstellar disks is similar to that in T Tauri stars. 
They propose that the lack of disks around HBe stars is due to the
rapid evolution of these objects, which disperse the surrounding dust and gas
in about 10$^6$ yrs \citep{fue02}. However, this lack of detection
can be a sensitivity effect. HBe stars are usually further away than HAe stars,
and
higher sensitivity interferometric observations  are required to
detect circumstellar disks around these objects. 
In this Letter, we present the first results of  a high-sensitivity search for circumstellar 
disks around three B0 stars, MWC~1080, MWC~137 and R~Mon,  which are the best studied of the five
Herbig B0 stars in the northern sky listed by \citet{the94}.

\section{Observations}
Interferometric observations in the continuum
at 1.4mm and 2.7mm have been carried out towards the Herbig Be stars 
MWC 1080, MWC 137 and R Mon, using the IRAM\footnote{IRAM is 
supported by INSU/CNRS (France), MPG (Germany) and IGN (Spain). } array at 
Plateau de Bure, France, in the CD set of configurations. 
MWC~137 and R~Mon, were also observed with the 
NRAO\footnote{The NRAO is a facility of the National Science Foundation operated
under cooperative agreement by Associated Universities, Inc.} Very Large Array
(VLA) at  0.7~cm and 1.3~cm in its D configuration.
Flux calibration is accurate within 10\% at 2.7mm and 20\% at
1.4mm in the PdBI images and within 20\% in the VLA images. No correction for primary 
beam attenuation has been applied. PdBI and VLA images are shown in
Fig~1. Flux densities at the star positions are shown in Table 1.

\begin{figure}
\includegraphics{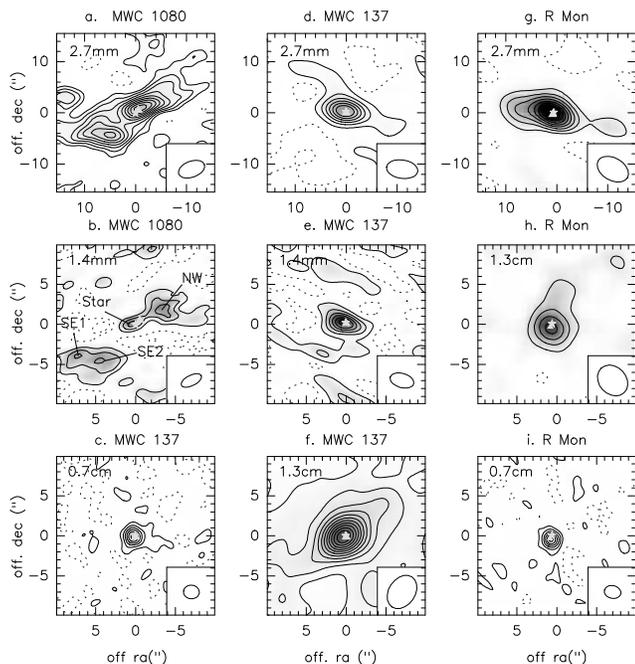}
\vspace{10cm}
\caption{Continuum images obtained with the PdB and VLA arrays.
The star marks the optical position [MWC 1080 : 23:17:25.574 60:50:43.34;
MWC 137: 06:18:45.504 +15:16:52.4; R Mon: 06:39:09.94 +08:44:10.0]
and the filled triangle the position 
of the radio source \citep{ski93}. Contour levels are: 
{\bf a.} -2.0 to -0.2 mJy/beam and 0.2 to 4.0 mJy/beam by  0.2 mJy/beam; 
{\bf b.} -1.6, -0.7, 0.7 to 3.4 mJy/beam by  0.9 mJy/beam;
{\bf c.} -0.2 mJy/beam, 0.2 to 2.2 mJy/beam by 0.4 mJy/beam; 
{\bf d.}  -0.3 mJy/beam,  0.3 to 3.6 mJy/beam by  0.6 mJy/beam;
{\bf e.}  -1.8 mJy/beam, -0.6 mJy/beam 0.6 to 6.6 mJy/beam by  1.2 mJy/beam; 
{\bf f.}  -0.2 mJy/beam, 0.2 to 2.4 mJy/beam by 0.2 mJy/beam; 
{\bf g.} -0.5 mJy/beam, 0.5 to 4.0 mJy/beam by 0.5 mJy/beam;
{\bf h.} -0.2 mJy/beam, 0.2 to 0.8 mJy/beam by 0.2 mJy/beam; 
{\bf i.} -0.2 mJy/beam 0.2 to 1.4 mJy/beam by 0.2 mJy/beam. \label{fig1}}
\end{figure}

\section{MWC 1080}
Although, we have not observed MWC~1080 at cm wavelengths.
some information about the cm emission 
can be found in the literature. \citet{cur89}, from 10$''$ resolution 
observations at 6cm, detected a weak extended component  which 
includes the star and the reflection nebula. 
\citet{ski93} with higher angular resolution ($\sim$ 1$''$)  tentatively 
detected at 3.6cm a compact source at the star position.
Recently, \citet{gir02} detected three 6cm sources using the VLA
with an angular resolution of $\sim$ 5$''$. 
One of these sources is  coincident
with the star position. They did not detect a counterpart of
this emission at 2.0cm with an upper limit of  $\leq$ 0.32 mJy. 
 
Intense emission is detected in the millimeter continuum images observed with
the PdBI (see Fig~1ab). Interferometric observations at 2.7mm
in MWC 1080 were previously reported by \citet{dif97} with a 
non-detection. Our image improves by a factor of 10 the sensitivity and by
a factor of 2 the angular resolution of their data.
The millimeter continuum emission in MWC~1080 is 
extended.  Two intense components connected
by a bridge of weak emission are observed in the 2.7mm image. These 
components are much closer to the star than the VLA 6cm sources detected by
\citet{gir02}.  We will refer to these components as $``$NW" and $``$SE". 
The better angular resolution of the 1.4mm image
resolve the $``$NW" and $``$SE" clumps in several components,  $``$NW1",$``$SE1",
$``$SE2" and $``Star"$. The total flux in the 1.4mm image is $\sim$ 24 mJy. In our single-dish
data we measured a 1.4mm flux of 73.8 mJy at the star position
with an angular resolution of $\sim$ 10$\farcs$5 \citep{fue98}. 
Thus the interferometer
has resolved the 1.4mm emission and missed $\sim$ 70\% of the flux.
The "Star" component is coincident with the optical star position
and remains unresolved in the 1.4mm image. This
implies that its size is $<$ 2$"$ which corresponds to a spatial extent of
$<$2000~UA at the distance of MWC 1080 (d=1000pc).   
This size is compatible with the emission arising in a circumstellar disk.
 
\section{MWC 137}
MWC~137 is associated with the one-arcminute size HII region S266 \citep{fic93}. 
Observations of this region at 3.6cm were reported by
\citet{ski93}. Our 1.3cm image of MWC~137 consists of a compact source 
coincident with the HBe star surrounded by a thin shell
with a radius of 30$"$-40$"$ from the point source. Weak emission is found towards
the thin shell and the point source. 
In the 0.7cm image,
the compact source appear more intense while the emission of the thin shell becomes
fainter. 
Since our aim is the detection of a circumstellar disk,
 we are only interested in the compact  component (see Table~1).

The continuum millimeter images show intense emission centered at the star position.
The total flux in the 1.4mm image is $\sim$ 10 mJy, i.e. $\sim$ 30\%
of the flux measured in the single-dish observations \citep{fue98}. 
From our 1.4mm image we derive a source size of 1.8$''$ $\times$ 0.8$''$,
which corresponds to  $\approx$ 2300~AU~$\times$~1000~AU assuming 
a distance of 1300~pc, and a position angle of 75$\pm$15$\deg$. Note that
the accuracy of the position angle estimate is heavily limited by the
S/N ratio of the data.

\section{R Mon} 
R~Mon was observed at 6cm, 3.6cm and 2cm using the VLA by \citet{ski93}.
They detected continuum emission at a position slightly shifted from the optical 
one. A compact source is detected in R Mon at 1.3cm and 0.7cm. The position of this 
source is offset (0.74$"$,-0.45$"$) from the optical position
but coincident with the radio source. A jet-like feature is observed in the 1.3cm image 
which is
surely driven by the radio source.  This jet was already detected by \citet{bru84}
in optical lines. Since the uncertainty of the optical position is not known,
the identification of the radio source with R Mon is uncertain. 

We have detected continuum emission towards the star in the 2.7mm image
with a peak emission of 4.1 mJy/beam. The  total integrated flux is 6.4 mJy
revealing that the emission is extended.  
Deconvolving the 2.7mm image with the beam, 
we estimate an emission size of 3$"$--4$"$, which
corresponds to  $\approx$ 3000~AU at the distance of 800 pc. 
\citet{nat00} reported a total 2.7mm flux of 13.0$\pm$1.3 mJy towards R Mon based 
on unpublished data by Mannings (1998). This flux exceeds by a factor of 2 our 
present result. This reinforces our conclusion that the emission is extended.
Then, the different beams and calibration uncertainity
can account for this discrepancy.The  2.7mm continuum emission is elongated in the direction 
perpendicular to the bipolar nebula axis and the centroid of the emission
lies to the northeast of the position of the radio-source.  
The shape and position of  the 2.7mm emission 
suggest that the 2.7mm emission is tracing a different emission component 
from that traced by the emission at cm wavelengths, and argues in favor of the existence
of a flattened structure (disk) around the star.

\begin{table}
%\begin{center}
\caption{Observed fluxes.\label{tbl-3}}
\begin{tabular}{l cc|cc} \hline       
\multicolumn{1}{l}{Object} &       
\multicolumn{1}{c}{S$_{1.3cm}$} &
\multicolumn{1}{c|}{S$_{0.7cm}$}  &
\multicolumn{1}{c}{S$_{2.7mm}$} &
\multicolumn{1}{c}{S$_{1.4mm}$}\\
\multicolumn{1}{l}{} &       
\multicolumn{1}{c}{(mJy/beam)} &
\multicolumn{1}{c|}{(mJy/beam)} &
\multicolumn{1}{c}{(mJy/beam)} &
\multicolumn{1}{c}{(mJy/beam)} \\
\hline
MWC 1080 & ...  & ... & $<$ 1.7 & 3.1 (0.2)  \\ 
MWC 137  & 1.5(0.5)$^1$  & 2.35(0.13) & 4.1(0.2) & 7.1(0.6)  \\
R Mon  &  0.78(0.10) & 1.26(0.13) & 4.1(0.5) & $<$ 15 \\
\hline
\end{tabular}

\noindent
$^1$ Subtracting the emission of the extended component
%\end{center}
\end{table}

\begin{figure*}[t]
\includegraphics{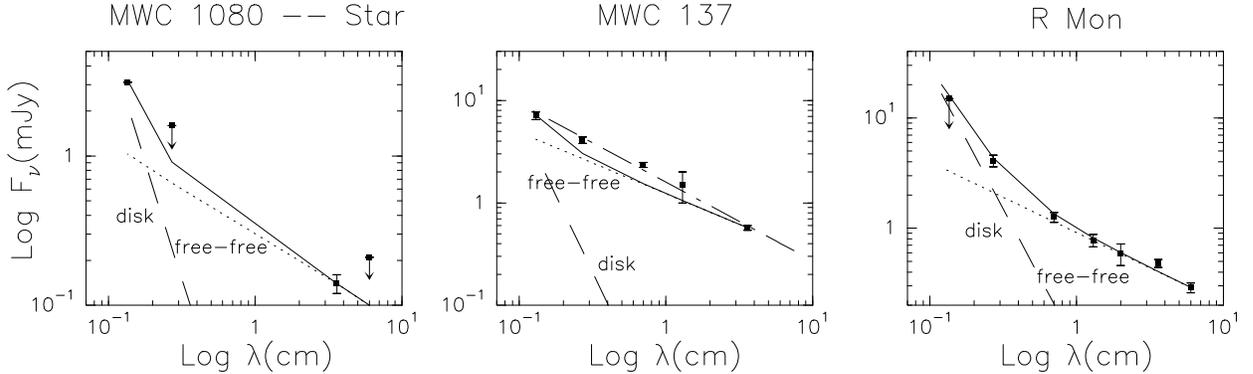}
\vspace{6cm}
\caption{ Spectral energy distributions (SEDs) at cm and mm wavelengths of 
MWC$\,$1080, MWC$\,$137 and R Mon. The fluxes at cm
wavelengths have been taken from Skinner et al. (1993) and Girart et al. (2002),
and at mm wavelengths are from this paper. 
The short-dash lines are model predictions for 
the free-free emission component. 
The long-dash lines are model predictions for
circumstellar disks of  0.003 M$_\odot$ and $\beta$=1 (MWC 1080), 
0.007 M$_\odot$ and $\beta$=1 (MWC 137),
and 0.01 M$_\odot$ and $\beta$=0.5  (R Mon). The solid line is the sum of both
components. The short-long straight line in central panel
is the predicted fluxes for the emission arising in a stellar wind with 
$\alpha$= +0.76 
in MWC 137. \label{fig2}}
\end{figure*}

\section{Spectral Energy Distributions (SEDs)}
The SEDs in the cm-mm range of MWC~1080,
MWC~137 and R~Mon are shown in Fig.~2. In order to
avoid the contribution of the envelopes of these sources
to the continuum emission,
only interferometric measurements have been included in
these SEDs. Even though, some problems still remain.
The emission at cm wavelengths 
in MWC~1080 is extended. The negative index derived by comparing the flux 
at 6cm and 3.6cm is very likely due to the larger beam of the 6cm flux
( 5.5$''$$\times$5.03$''$) compared with that of the 3.6cm one (1.24$''$$\times$0.83$''$).
The disk emission should be unresolved at the distance of our source.
Therefore, we are not interested in the extended
component and consider the 6cm flux as an upper limit. 
Unfortunately, the free-free emission arising in the stellar wind is also 
expected to be unresolved with our beam
and could make an important contribution to the continuum flux even at millimeter 
wavelengths. In order to derive the disk mass in MWC 1080, we have 
assumed that the 3.6cm flux is due to free-free wind emission and subtracted
it from the observed mm fluxes. These corrected values are then
interpreted as optically thin disk emission, according to the expression
$F_{\lambda} = d^{-2} M_d \kappa_{\lambda} B_{\lambda}(T_d)$
with $\kappa = 0.01 ( \lambda (mm) / 1.3 )^{-\beta} cm^{-2} g^{-1}$.
We have assumed a constant value of 
$T_d$ = 215 K following the recommendation of
\citet{nat00} for B0 stars.
With these assumptions, we obtain from the 1.4mm flux a disk mass of 0.003 M$_\odot$. 
Since we have only an upper limit to the 2.7mm flux, the value of $\beta$ cannot
be determined. 
We have compared the disk mass derived with 
this simplified expression to those of more elaborated  models  (see
Dullemond et al. 2002), varying the disk outer radius and the surface density profile
over a large range of possible values, and confirmed the conclusion of
\citet{nat00} 
that the values of  M$_d$ derived in this way are
accurate  within a factor  2 -- 3. 
 We have also considered the possibility that the 3.6cm emission
arise in an optically thin HII region instead of free-free emission.
With only two points in the SED, we cannot discern
between these two possibilities. In this case, we obtain a slightly larger value of
the disk mass of, M$_d$ $\sim$ 0.004 M$_\odot$. 
Therefore we estimate that the mass of the disk around MWC~1080 
is 0.003$\pm$0.001 M$_\odot$.

The complete SED of MWC~137 can be fitted
with a single component with spectral index $\alpha$ = +0.76$\pm$0.01. 
This spectral index is consistent with that expected in the free-free emission 
arising in the stellar wind. Although a value of $\alpha$ =+0.6 is expected
for a ionized isotropic wind, small deviations of this value can be explained by
a different geometry or a partially ionized wind.  However, we cannot discard
the existence of a very low-mass disk.
In order to estimate an upper limit to the disk mass, we have extrapolated the
3.6cm flux to millimeter wavelengths with an spectral index of +0.6. Then,
we have estimated the disk mass from the excess flux at 1.4mm. With the same
values of $\kappa$ and T$_d$  than in the case of MWC 1080, we obtained an
upper limit to the disk mass of 0.007 M$_\odot$. 

The SED of R~Mon is the most complete one. We have been able to fit all the
centimeter  data using a spectral index, 
$\alpha_{cm}$ $\approx$ 0.64, which
is characteristic of a spherical stellar wind. In order to estimate the mass of the
circumstellar disk, we have extrapolated the free-free emission to millimeter 
wavelengths and fit the millimeter part of the SED with a simple disk model.
As in the case of MWC 1080, we have assumed an unresolved disk at a 
constant temperature, T$_d$ = 215 K and adjusted the value of M$_d$ and
$\beta$. Only values of $\beta$ = 0.25 -- 0.5 fit our observational points.
A value of $\beta$ larger than 0.5 would produce an excess of flux at 1.4mm. 
This low value of $\beta$ could be interpretred as an optically thick
compact disk, or an optically thin disk with large grains. The first possibility 
would imply millimiter fluxes larger than those measured. A small disk of R = 100 AU
should have a mass of 0.35 M$_\odot$ in order to have  $\tau_{1.4mm}$$\sim$1.
But in this case, the 1.4mm flux would be at least an order of magnitude
larger than our upper limit. The situation would be worse
for a larger disk.
The second possibility seems more plausible.
Similarly low values of $\beta$ have been found in Herbig Ae and T Tauri
disks and reveal highly processed dust \citep{bec91,tes03}. 
With the fiducial value  of the dust emissivity, we  derive a disk mass of 0.01 M$_\odot$. 
Note, however, that, if the low value of $\beta$ is interpreted as evidence of grain growth, 
one needs to reduce the fiducial 2.7mm dust emissivity by a factor of 
about 4 \citep{tes03},
and  to increase the estimated disk mass by a similar amount.

\begin{figure}[t]
\includegraphics{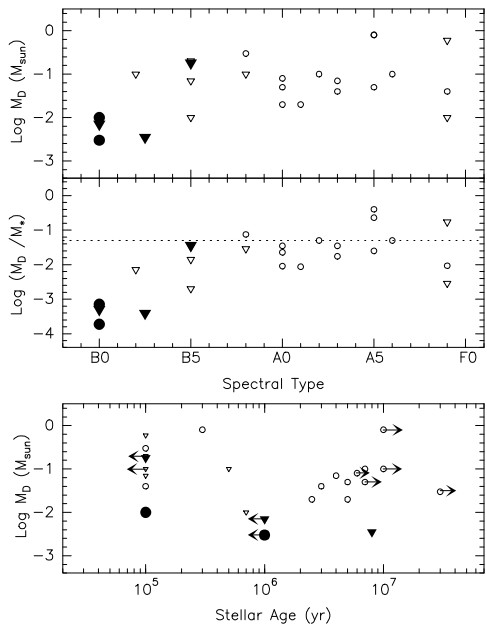}
\vspace{9cm}
\caption{Plot of the M$_d$ and M$_d$/M$_*$ versus the spectral type 
of the star ({\bf top panels}) and M$_d$ versus the stellar age
({\bf bottom panel}) for a wide sample of HAEBE stars.
Data have been taken from the compilation 
by \citet{nat00} (empty symbols) and the results reported by \citet{fue01} and this paper 
(filled symbols).Dots mean detections while triangles mean upper limits. Note that
disk masses in HBe stars are at least an order of magnitude lower than in HAe stars.
 \label{fig3}}
\end{figure}

\section{Disk detections in HBe stars}
Our results prove the existence of circumstellar dust around 
MWC 1080 and R Mon. However, it is very difficult to interpret it
beyond doubt as evidence
of a circumstellar disk. In general,
a compact and small 1.4mm source  ($<$ 1$''$ -- 2$''$) at the star position is 
considered as a disk detection. This criterium seems sufficient 
in TTS located at distances of $\sim$ 150 pc, in which an 
angular resolution of 1$''$ corresponds to a linear scale of 100 - 200 AU ,
typical size of circumstellar disks.
Since HBe stars are usually located at distances larger than
1 Kpc, even the higher angular resolution provided by interferometers,
$\sim$ 1$''$, corresponds to sizes of a few thousands of AU, and
it is not possible to distinguish between ``bona-fide"
circumstellar disks and
a flattened structure of a few 1000~AU. 

Another difficulty comes from the fact that many HBe stars are likely associated 
to lower mass companions that will not be resolved in our beam.
MWC 1080 is a close binary with a separation of 0.7 $''$ \citep{lei97}.
Mid-infrared observations by \citet{pol02} using the Keck II telescope
resolve the binary and found that  warm material ($\sim$ 350 K),
with a  3 -- 10 $\mu m$ spectral index consistent with the existence of a 
circumstellar disk, is surrounding  MWC 1080 A. A cooler dust component
is  associated with MWC 1080 B.  
We have detected an unresolved mm continuum source in MWC 1080. We
consider that this source is a good candidate for disk detection
around a HBe star. However,  since the angular resolution of our 
observations is not sufficient to resolve
the two binary components, we cannot exclude the possibility that the
emission is associated with the younger companion.

R Mon is also a close binary with a separation of 0.69$"$.
Its companion  is a very young TTS. 
\citet{clo97} found 
an extinction of Av = 13.1 mag towards the star which 
they interpreted as due to an optically thick disk of R = 100 AU.
Our 2.7mm image reveals the existence of a compact source
of  $\approx$ 3000 AU $\times$ 2300 AU. This size is larger than that
expected for a circumstellar disk and suggests the existence
of a flattened envelope surrounding the disk.  
Our value of the disk mass may be overestimated due to the
contribution of the flattened envelope. However, the
analysis of the mm spectral index shows
that the grains responsible of the emission are 
heavily processed, so that 
most of the emission is very likely arising
in a circumstellar disk.  
If this is true, it is unlikely that the disk could be associated to the
TTS companion, since in this case one should use in deriving 
the disk mass a value of $T_d \sim 15$ K \citep{nat00}
and the estimated mass would increase
to values in the range 0.2--0.5 M$_\odot$, too large for a TTS disk.

We have completed the compilation of (interferometric)
disk observations of HAeBe stars carried out by \citet{nat00}
with the observations in MWC 1080, MWC 137 and R Mon
reported in this paper, and the observations in HD 200775 and
LkH$\alpha$ 234 reported by \citet{fue01}. In Fig. 3, 
we plot the disk masses as a function of the spectral type of the 
star  and the stellar age. We have plotted
MWC 1080 and R Mon as disk detections; however, as discussed in the previous
section, it is possible that the ``disk'' emission is either overestimated
or not associated to the Herbig Be star itself,
but to unresolved companions, thus  reinforcing our conclusions.
In any case, disk masses in HBe stars are at least an order
of magnitude lower than in HAe stars. Plotting M$_d$/M$_*$ as a function 
of the spectral type, one finds that
while the value of M$_d$/M$_*$ is roughly constant and equal to 0.04  for
stars with spectral type A0-M7, M$_d$/M$_*$ $<$ 0.001 in
HBe stars.    
The same effect remains when one plots the disk mass against
the stellar age. The values of disk masses in HBe stars are systematically lower
than in HAe stars for stars with ages between 10$^5$ and 10$^7$ yrs.
This lack of massive disks in HBe stars even for stars as young as
10$^5$ years put several constraints to the possibility of forming
planetary systems around massive stars. In a time-scale of $<$ 10$^6$ years
the disk mass around a HBe star is similar to or even lower than
the total 
mass of the  planets in the solar system ($\sim$ 0.0013 M$_\odot$). Thus 
planet formation should occur very fast, for 
planetary systems 
to exist at all around  massive stars. 

\acknowledgments
 This paper has been partially funded by the Spanish MCyT 
  under projects DGES/AYA2000-927, ESP2001-4519-PE 
  and ESP2002-01693, and European FEDER funds. 
A.N. and L.T. acknowledge support from ARS-1/R/073/01 grant to the 
Osservatorio di Arcetri. A.R. acknowledge support from ES P2002-01627 y
AYA2002-10113-E





\end{document}